\documentclass[10pt,conference]{IEEEtran}
\usepackage{amsmath,amssymb,amsfonts}
\usepackage{graphicx}
\usepackage{textcomp}
\usepackage{xcolor}
\usepackage{comment}
\usepackage{multirow}
\usepackage{hyperref}
\hypersetup{
    colorlinks=true,
    linkcolor=black,
    citecolor=black,
    urlcolor=blue
    }
\def\BibTeX{{\rm B\kern-.05em{\sc i\kern-.025em b}\kern-.08em
    T\kern-.1667em\lower.7ex\hbox{E}\kern-.125emX}}
    
\newcommand{\tabincell}[2]{\begin{tabular}{@{}#1@{}}#2\end{tabular}}

\newcommand*{\img}[1]{%
	\raisebox{-.3\baselineskip}{%
		\includegraphics[
		height=\baselineskip,
		width=\baselineskip,
		keepaspectratio,
		]{#1}%
	}%
}

\makeatletter
\newcommand{\linebreakand}{%
  \end{@IEEEauthorhalign}
  \hfill\mbox{}\par
  \mbox{}\hfill\begin{@IEEEauthorhalign}
}
\makeatother

\begin{document}

\title{SeeAction: Towards Reverse Engineering How-What-Where of HCI Actions from Screencasts for UI Automation}

\author{\IEEEauthorblockN{1\textsuperscript{st} Dehai Zhao}
\IEEEauthorblockA{
\textit{CSIRO's Data61}\\
Sydney, Australia \\
dehai.zhao@data61.csiro.au}
\and
\IEEEauthorblockN{2\textsuperscript{nd} Zhenchang Xing}
\IEEEauthorblockA{
\textit{CSIRO's Data61 \& School of Computing, ANU}\\
Sydney, Australia \\
zhenchang.xing@data61.csiro.au}
\and
\IEEEauthorblockN{3\textsuperscript{rd} Qinghua Lu}
\IEEEauthorblockA{
\textit{CSIRO's Data61}\\
Sydney, Australia \\
qinghua.lu@data61.csiro.au}
\linebreakand
\IEEEauthorblockN{4\textsuperscript{th} Xiwei Xu}
\IEEEauthorblockA{
\textit{CSIRO's Data61}\\
Sydney, Australia \\
xiwei.xu@data61.csiro.au}
\and
\IEEEauthorblockN{5\textsuperscript{th} Liming Zhu}
\IEEEauthorblockA{
\textit{CSIRO's Data61 \& School of CSE, UNSW}\\
Sydney, Australia \\
liming.zhu@data61.csiro.au}
}

\maketitle

\begin{abstract}
UI automation is an useful technique for UI testing, bug reproduction and robotic process automation.
Recording the user actions with an application assists rapid development of UI automation scripts, but existing recording techniques are intrusive, rely on OS or GUI framework accessibility support or assume specific app implementations.
Reverse engineering user actions from screencasts is non-intrusive, but a key reverse-engineering step is currently missing - recognize human-understandable structured user actions (\textcolor{red}{[command]} \textcolor{cyan}{[widget]} \textcolor{green}{[location]}) from action screencasts.
To fill the gap, we propose a deep learning based computer vision model which can recognize 11 commands and 11 widgets, and generate location phrases from action screencasts, through joint learning and multi-task learning.
We label a large dataset with 7260 video-action pairs, which record the user interactions with Word, Zoom, Firefox, Photoshop and Windows 10 Settings.
Through extensive experiments, we confirm the effectiveness and generality of our model, and demonstrate the usefulness of a screencast-to-action-script tool built upon our model for bug reproduction.
\end{abstract}


\section{Introduction}
\label{sec:introduction}

UI automation reduces or assists the repetitive, manual digital tasks of a human worker by a ``digital worker''. 
In the context of software engineering, it is used to simulate an end-user's interaction with an application for UI testing~\cite{guo2019sara, sahin2019randr, qin2016mobiplay, yu2021layout} or bug reproduction~\cite{chaparro2019assessing, zhao2019recdroid, moran2016automatically, fazzini2018automatically}.
Beyond software engineering, UI automation is a core capability in robotic process automation~\cite{qian2020roscript} which aims to automate business operations such as invoice processing, inventory or human resources management. 
UI automation can be achieved through textual-script or visual-script based techniques, such as Selenium~\cite{selenium}, Appium~\cite{appium}, UIAutomator~\cite{uiautomator}, UiPath~\cite{uipath} and AirTest~\cite{airtest}.
UI Automation scripts can be coded from scratch, which involves repetitive and error prone manual work.
In practice, UI automation tools often support record-and-replay for recording human-computer interaction (HCI) actions from which automation scripts can be produced.
The recorded scripts can be automatically replayed or be used as a boilerplate script to expand from. 

\begin{figure}
	\centering
	\includegraphics[width=0.95\linewidth]{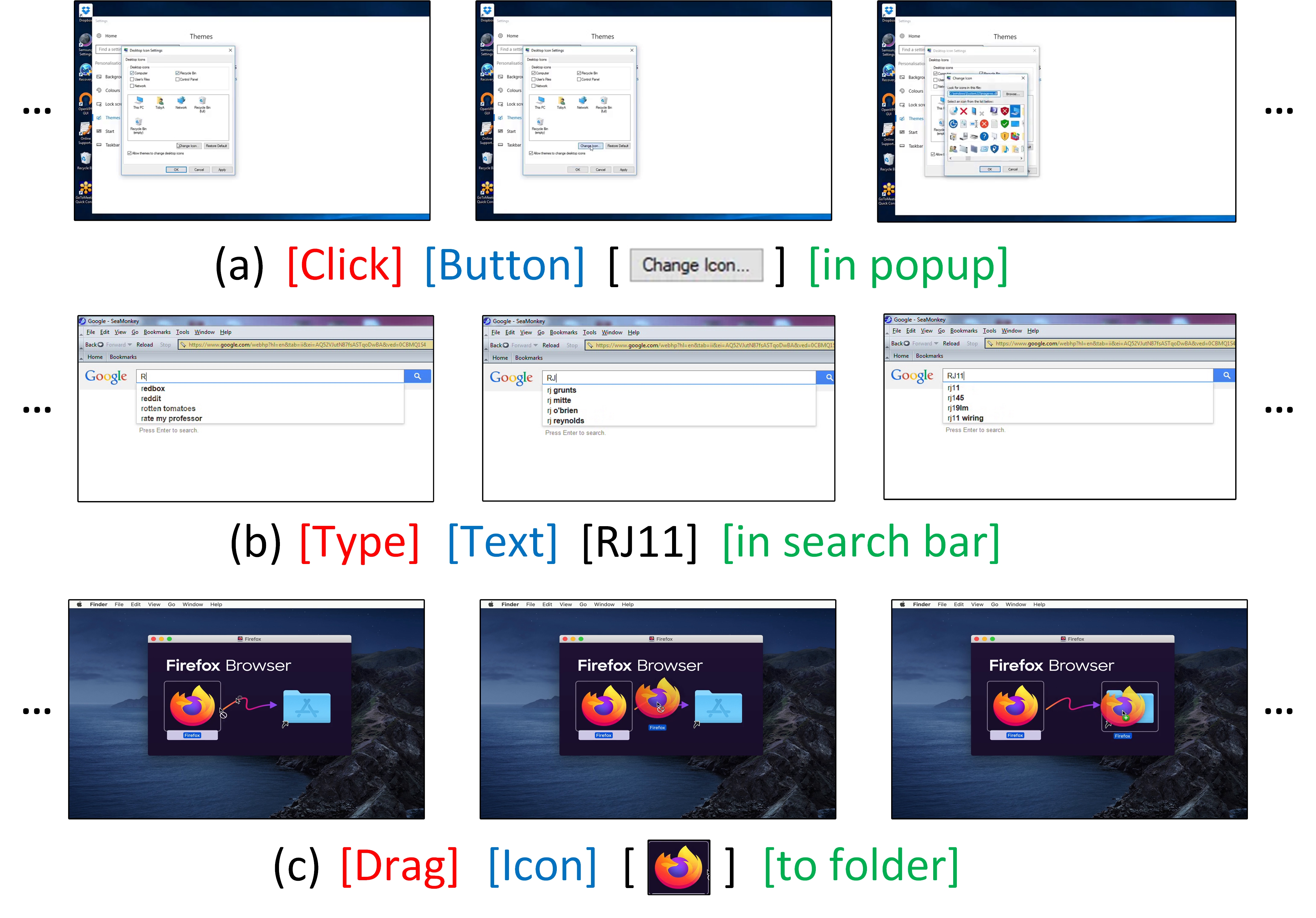}
	\caption{Examples of structured HCI actions}
	\label{fig:structured}
\end{figure}

HCI actions can be recorded at three levels: operating system (OS)~\cite{gomez2013reran, halpern2015mosaic, bao2015activityspace, bao2018vt}, GUI framework~\cite{hu2015versatile} or application~\cite{qin2016mobiplay, espresso, appium, selenium, guo2019sara}.
Recording at application level is preferable because it outputs human understandable scripts, rather than system or API calls.
Figure~\ref{fig:structured} shows some examples of application-level HCI actions which we refer to as \textit{structured HCI actions} in this work.
Structured refers to three integral parts of an action: 
\textit{how} (command, e.g., click, type, drag); 
\textit{what} (widgets, if any, that the action operates, e.g., button, text, icon); and
\textit{where} (location in the UI where the action occurs, e.g., popup, search bar, folder).
Such structured HCI actions are in a human understandable format that we see in application usage tutorials~\cite{wikihow} and steps-to-reproduce descriptions in bug reports~\cite{zhao2019recdroid, moran2016automatically, fazzini2018automatically}.
Furthermore, they are robust to changes of screen resolutions and sizes, and changes in spatial arrangements of GUI widgets, compared with low-level screen coordinates.

Some HCI recording tools require app source code (e.g., Espresso~\cite{espresso}) or customized OS (e.g., VALERA~\cite{hu2015versatile}). 
Due to their intrusiveness, such recoding tools have limited applicability~\cite{lam2017record}. 
Most recording techniques are less intrusive, but they still require either accessibility support from OS or GUI framework~\cite{appium, uiautomator} or dynamic instrumentation of software applications~\cite{guo2019sara, sahin2019randr}.
As such, user actions involving non-accessible-friendly widgets, for example custom widgets like non-Android webview in hybrid apps, cannot be effectively recorded.
Dynamic instrumentation hooks to specific APIs (e.g., event dispatcher, Android TextView for rich text editor) and is sensitive to system and implementation changes.
It may also introduce the incompatibility with the instrumented apps~\cite{lam2017record}.


It is desirable to develop non-intrusive record-and-replay techniques, independent of OS, GUI frameworks and application implementations.
This non-intrusiveness is important to support user experience testing~\cite{amobile} and to support closed embedded systems that are increasingly used in entertainment, education and industry~\cite{qian2020roscript}.
Screen recording non-intrusively produces a video recording of HCI actions.
Structured user actions can be non-intrusively reverse-engineered from screen recordings using computer vision techniques.
This reverse-engineering process involves three steps:
first, \textit{video segmentation} to segment the video into video fragments of individual actions;
second, \textit{structured action recognition} to generate how-what-where information of HCI actions from video fragments;
third, \textit{widget identification} to identify GUI widget details involved the actions.
Techniques have been developed for video segmentation~\cite{xu2017r, zhao2017temporal}, simple HCI action prediction~\cite{zhao2019actionnet,bernal2020translating}, and widget detection~\cite{xie2020uied}.
However, effective techniques for structured action recognition from action videos are missing, without which we will not have a working reverse-engineering pipeline.

Recently, large language models (LLMs) have gained popularity, with some advancing to support vision features\cite{chatgpt}. 
However, supporting video action recognition remains a significant challenge. 
While some work has developed LLMs for action recognition in videos\cite{zhao2023learning, bi2023misar}, these models primarily support videos in natural scenes. 
It has been demonstrated that screencasts present different challenges compared to natural scenes\cite{zhao2019actionnet}. 
Moreover, LLMs require substantial computational resources. 
It is not necessary to employ large and heavy models for all problems. 
A specialized and compact AI model is a more efficient and effective choice for addressing domain-specific problems.

This work fills the gap.
We propose a novel model to recognize user actions in screencast fragments and generate structured natural language descriptions of user actions such as those in Figure~\ref{fig:structured}.
Given an action screencast, we first compute the structural similarity~\cite{wang2006modern} of adjacent frames to obtain change regions and similarity maps.
Our model takes three data streams as input: the sequences of original video frames, cropped change regions and similarity maps.
Then the three data streams pass through three 3D Covolutional Neural Networks (CNNs)~\cite{tran2015learning} respectively which extract spatio-temporal features.
The features from three channels are concatenated for three tasks: command classification, widget classification and location phrase generation.
We define 11 classes of commands and 11 classes of widgets to be recognized from screencast fragments, based on previous literature~\cite{bao2015activityspace, banovic2012waken, zhao2019actionnet} and the actions and widgets supported by visual script based automation tools such as Airtest~\cite{airtest} and sikulix~\cite{sikulix}.
Location phrase is free form that covers common expressions of locations on UI from application usage tutorials~\cite{wikihow}.
The model is trained end-to-end by joint learning and multi-task learning.

To train and evaluate our model, we build the first dataset of screencasts labeled with structured user action descriptions, which contains a total of 7,260 video-action pairs.
The videos are application demonstration videos from YouTube and bug reproduction screencasts from Firefox.
We select five desktop applications with distinct functionalities: Word, Zoom, Firefox, Photoshop, Windows 10 Settings.
The applications run on Windows, Linux and macOS.
The videos are segmented and labeled with structured user-action descriptions manually by the two authors.
Our experiments on this dataset show that our model achieves high recognition accuracy (0.82 F1-score for command classification, 0.85 F1-score for widget classification, and 0.77 BLEU~\cite{papineni2002bleu}, 0.51 METEOR~\cite{banerjee2005meteor}, 0.76 ROUGE~\cite{lin2004rouge} and 2.85 CIDEr~\cite{vedantam2015cider} for location phrase generation).
Furthermore,  the joint learning of the three data streams can significantly improve the model performance, and the multi-task learning can also enhance the model performance, especially for location phrase generation.

To demonstrate the usefulness of our model, we integrate it with a simple heuristic-based video fragmentation method and the GUI widget detection method~\cite{xie2020uied} into a screencast-to-actionscript tool.
We apply this tool to 100 bug reproduction screencasts from Firefox and obtain the action scripts for reproducing these bugs.
In a comparative study, the Firefox users reproduce 45\% more bugs following the action scripts generated by our tool, compared with those following the original steps-to-reproduce (S2R) descriptions in the bug reports due to many S2R quality issues such as ambiguous steps, vocabulary mismatch, missing steps, or even absent S2R text~\cite{chaparro2019assessing}.
The results show the potentials of our model for supporting UI automation tasks such as bug reproduction.

We make the following contributions in this paper:
\begin{itemize}
	\item To the best of our knowledge, this is the first work to recognize structured user actions from screencasts. It is an essential step towards non-intuitive UI automation.
	
	\item To recognize structured user actions from screencasts, we propose a novel computer vision model, which is trained end-to-end by joint learning and multi-task learning.
	
	\item We label and contribute the first screencast dataset with 7260 video-action pairs, which record the user interactions with five popular desktop applications.
	
	\item Through extensive experiments, we not only confirm the effectiveness and generality of our model, but also demonstrate its usefulness for generating high-quality, human-understandable action scripts for bug reproduction.
\end{itemize}

\section{Approach}
\label{sec:approach}
Given a screencast fragment recording of a user action with an application, our approach outputs a structured natural language description of this action.
The screencast fragment is first processed by computer vision techniques to obtain three data streams, i.e., the sequence of original frames, cropped change regions and similarity maps.
Next, the three image sequences pass through three 3D CNNs~\cite{tran2015learning} respectively to extract spatio-temporal features (see Figure~\ref{fig:model}), which are finally used to predict structured user actions including command category, widget category and location phrase.
We formulate the spatio-temporal feature extraction as joint learning and structured user action prediction as multi-task learning.

\subsection{Structured HCI Actions}
\label{sec:define}

We summarize 11 command classes and 11 widget classes based on our personal experiences with software applications, previous work~\cite{zhao2019actionnet, bao2015activityspace, bao2018vt}, and the commands and widgets supported by popular visual-script automation tools (e.g., AirTest~\cite{airtest}, sikuli~\cite{sikulix}).
The 11 command categories include Click, Drag, Hover, Scroll down, Scroll up, Select, Type, Zoom in, Zoom out, Appear and Disappear.
The first nine commands are user operations, while Appear and Disappear are app responses (i.e., the outcomes of user operations).
We include app responses because they represent critical information in the screencasts for downstream applications, for example, assertions for UI testing~\cite{hu2015versatile}.
The 11 widget classes are Button, Checkbox (including Radio Button), Dropdown, Icon, Image, Text (including Text Field), Window, Page, Tab, Popup and Others.
The first six widget classes are atomic widgets for presenting information and collecting user inputs, while Window, Page, Tab and Popup are container widgets for grouping visual content.
We include others to represent miscellaneous widgets that are not widely present in applications (e.g., geometry graph).
Different from commands and widgets, the location information requires a free-form short phrase to express, which will be assembled from a vocabulary.
The vocabulary can be built from application usage tutorials (e.g., WikiHow~\cite{wikihow}) or from the labeled screencast dataset (see Section~\ref{sec:dataset}).
Note that command, widget and location have latent associations.
For example, user actions can be [click] [button] or [type] [text] [in search bar], but not [type] [button] or [click] [text] [in toolbar].

\subsection{Model Input}
A user interaction with application is recorded as a screencast.
The raw screencast will be processed into a fixed-length sequence of adjacently-distinct screenshots.
From this sequence of screenshots (frames), a grayscale similarity map is computed between adjacent screenshots and change regions are subsequently detected from the similarity map.
The sequence of the original screenshots, together with the obtained sequences of similarity maps and change regions, will be fed into the deep learning model.

\subsubsection{Input Screencast Processing}
\label{sec:videoprocessing}
We decode a raw screencast video into a sequence of frames at a sample rate of 5 frames per second (fps).
According to our empirical observation, 5fps sample rate keeps sufficient frames to represent HCI actions and screen changes, without too many redundant frames.
We further scan the frame sequence from the beginning and remove the adjacently-identical frames (i.e., no screen changes).
User actions vary in duration (0.2 to 10 seconds in our labeled data), leading to different frame sequence lengths.
But the deep learning model requires a fixed length sequence of screenshots.
We normalize an input frame sequence to a fixed length $S$ by down- or up-sampling.
Given a frame sequence $F = \{f_1, f_2, \dots, f_M\}$, we pick the first frame $f_1$ and the last frame $f_M$ to a sampled frame sequence $F' = \{f_1, f_M\}$, as these two frames contain the most important information to represent how a user action starts and ends.
Next, if $S \leq M$, we randomly down-sample $S - 2$ frames $\{f_i, \dots, f_j\}$ from the remaining frame sequence $F = \{f_{2}, \dots, f_{M - 1}\}$ and insert them into the sampled frame sequence $F' = \{f_1, f_i, \dots, f_j, f_M\}$ by the original frame order.
If $S > M$, we randomly duplicate the frames from the original frame sequence $S - M$ times and insert the obtained frames into the sampled frame sequence by the original frame order.
In this work, we set $S = 8$ as this length is long enough to cover the duration of most user actions and contains the least duplicate frames.

\subsubsection{Model Input Representation}
\label{sec:inputrepresentation}

Many user actions (e.g., click button, type text) result in only small-scale screen changes.
Image features in such small-scale screen changes would be too weak to recognize user actions if the whole frame is taken as the only input.
Therefore, we detect change regions which will be used as an additional input to the model.
First, we compute the structural similarity~\cite{wang2006modern} which produces a similarity score for each pair of pixels at the same location in the two frames.
Based on these similarity scores, we produce a grayscale image (i.e., a similarity map) in which brighter pixels indicate lower similarities between the corresponding pixels in the two frames.
On this similarity map, we compute the regions of connected non-black pixels and determine their bounding boxes.
The corresponding areas in original frames are cropped by these bounding boxes as change regions.
There can be more than one change region between the two frames and we keep the largest change region.
We prepare three input data streams:
1) the sequence of original frames (each frame is a RGB image) which provide the detailed context of user actions;
2) the sequence of cropped change-regions (each cropped change-region is a RGB image) which provide screen-change details resulting from user actions; and
3) the sequence of similarity maps (each map is a grayscale image) which provide a simplified overview of screen changes.
All input images are resized to the same size (e.g., $224 \times 224$).

\begin{figure*}
	\centering
	\includegraphics[width=\textwidth]{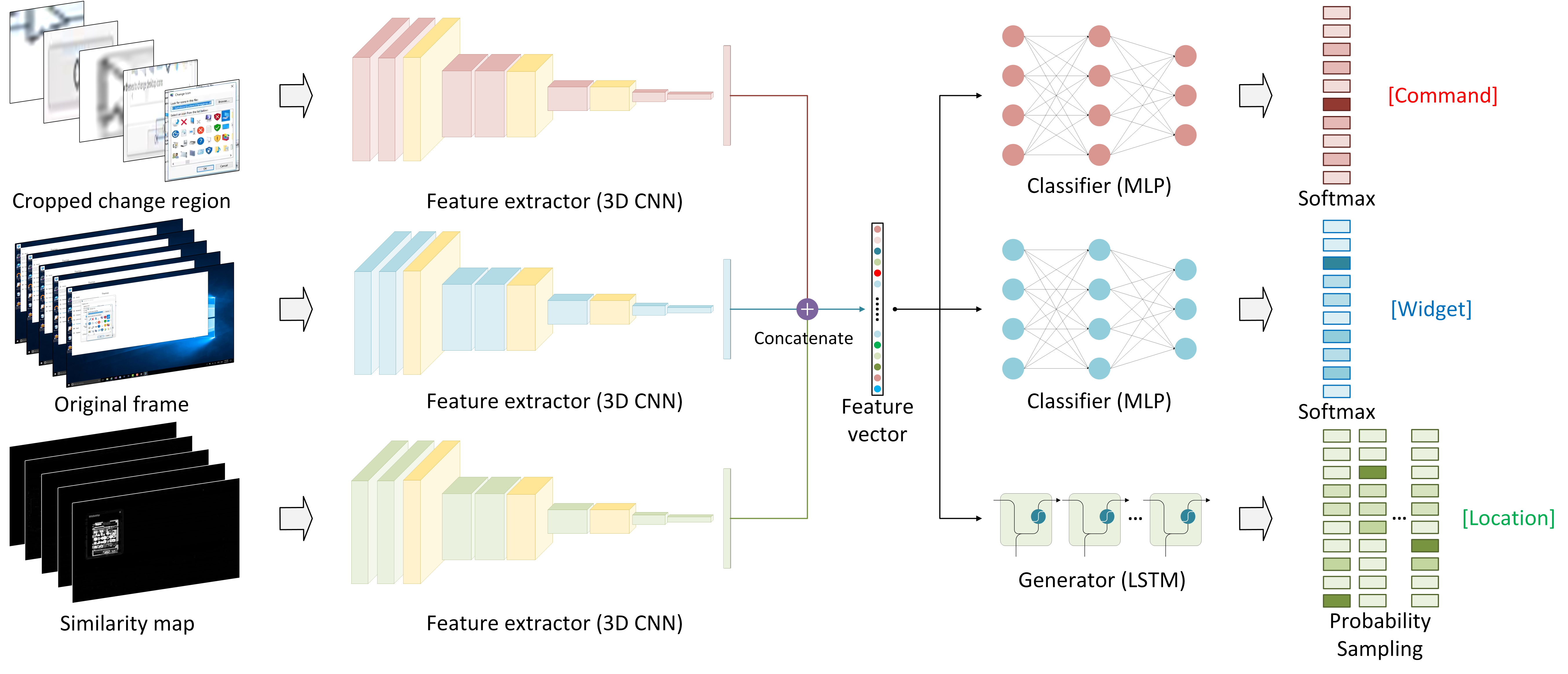}
	\caption{Overall model architecture of SeeAction}
	\label{fig:model}
\end{figure*}

\subsection{Model Architecture}

Our model takes as input three image sequences (original frames, change regions and similarity maps) obtained from a user-action screencast, and outputs a structured natural language description of the user action performed in the input screencast, in the form of [command] [widget] [location], such as [click] [button] [in popup].

\subsubsection{3D-CNN based Spatio-Temporal Feature Extraction}
User actions result in changes on screen and last for a short time period.
3D-CNNs has been successfully applied to video data for both natural scenes~\cite{soomro2012dataset} and computer rendered screencasts~\cite{zhao2020seenomaly} to encode video data into low-dimensional abstract spatio-temporal features for action recognition.

As shown in Figure~\ref{fig:model}, we use three different 3D-CNN feature extractors to encode the spatio-temporal features in the three input image sequences respectively.
The three 3D-CNN encoders share the same model structure but do not share weights.
In our model, we aggregate three 3D-CNNs as one spatio-temporal feature extractor and optimize them jointly.
Joint learning makes the whole model smaller and improves the efficiency of training process.
The three 3D-CNN encoders extracts a low-dimensional feature vector $E_{chreg}$, $E_{orig}$, $E_{simmap}$ for cropped change regions, original frames and similarity maps, respectively.
The model concatenates the three vectors $E = E_{chreg} \bigoplus E_{orig} \bigoplus E_{simmap}$ as an overall representation of the user action occurred in the action screencast.
We set the length of three feature vectors as 512, resulting in the length of fused feature vector $E$ as 1536.
The fused feature vector is fed into the subsequent multi-task action prediction.

\subsubsection{Multi-task User Action Prediction}
Considering the latent associations among [command] [widget] [location], we formulate the model prediction as three sub-tasks in a multi-task learning setting.
As command and widget comes from a fixed set of labels, we formulate command and widget prediction as two classification tasks, i.e., video tagging~\cite{soomro2012ucf101}.
As the location is a phrase assembled from a vocabulary, we formulate location prediction as a sequence-to-sequence generation task, i.e. video captioning~\cite{chen2015microsoft}.

The two classifiers and one generator produce three outputs, i.e., command class, widget class and location phrase, respectively.
We compute three loss functions for the three tasks.
The command classification and widget classification task use cross-entropy loss, which is denoted as $L_{act} = - \sum_{c = 1}^{N_{act}} y_c \log (p_c)$ and $L_{obj} = - \sum_{c = 1}^{N_{obj}} y_c \log (p_c)$ respectively, where $y$ is the ground-truth label and $p$ is the predicted class.
The location phrase generation task applies cross-entropy loss to each word output by LSTM, and it is computed as $L_{loc} = - \sum_{i = 1}^{maxL} \sum_{c = 1}^{N_{voc}} y^i_c \log (p^i_c)$, where $maxL$ is the maximum length of the generated text sequence, $y^i$ is the ground-truth label and $p^i$ is the predicted word at i-th position.
We formulate the three tasks as multi-task learning by computing the sum of three losses and get the total loss, i.e., $L = L_{act} + L_{obj} + L_{loc}$.
The training objective is to minimize the total loss $L$ and make the model reach global optimization.

\section{Experiment Design}
\label{sec:experiment}

This section describes our experiment dataset for model training and testing, and the metrics used for evaluation.

\subsection{Dataset}
\label{sec:dataset}
In this study, we manually label 288 application usage screencasts (in total 12.8 hours) for five desktop applications (see Table~\ref{tab:screencast}).
This creates a dataset of 7,260 screencast fragments labeled with corresponding structured user action descriptions, which to the best of our knowledge is the first large-scale dataset of its kind.

\subsubsection{Data collection}
We consider five popular desktop applications that most people use daily for diverse tasks, including Mozilla Firefox (web browsing), Microsoft Word (document processing), Adobe Photoshop (graphics editing), Zoom (online conference) and Windows 10 Settings (system configuration).
We collect application usage screencasts from two sources.
First, we collect application demonstration videos from YouTube by searching keywords ``How to use applicationname'' and download the most viewed ones at high definition resolution (1280$\times$720).
We keep the videos in which users are actually using the applications but exclude those recorded in slide presentation.
Second, we crawl steps-to-reproduce (S2R) screencasts from the Firefox's bug reports~\cite{Bugzilla}.
These steps-to-reproduce screencasts have diverse screen sizes and resolutions.
As shown in Table~\ref{tab:screencast}, we finally collect a total of 288 videos with 12.8 hours duration.
The collected videos involve three main stream operating systems (i.e. Windows, Linux and macOS).
Because of different usage characteristics of applications (e.g., the density of user actions), video numbers and duration vary among the five applications.
Zoom window may contain other application views, for example, screen shared by online conference participants, which poses an interesting challenge to predicting user actions with Zoom.

\begin{table}
	\centering
	\caption{Dataset of use-action screencasts}
	\begin{tabular}{|c|c|c|c|c|}
		\hline
		\textbf{Application} &\textbf{\#Video} & \textbf{Dur.(h)} & \textbf{OS} & \textbf{Resolution} \\
		\hline
		Photoshop      & 6   & 0.75  & macOS                            & 1280 $\times$ 720 \\
		\hline
		Win10 Settings & 21  & 3.83  & Windows                          & 1280 $\times$ 720 \\
		\hline
		Firefox        & 158 & 1.32  & \tabincell{|c|}{Linux \& \\ macOS} & Multiple \\
		\hline
		Zoom           & 55  & 4.70  & Windows                          & 1280 $\times$ 720 \\
		\hline
		Word           & 48  & 2.21  & Windows                          & 1280 $\times$ 720 \\
		\hline
		\textbf{Total}          & \textbf{288} & \textbf{12.82} & \textbf{-}                               & \textbf{-} \\
		\hline
	\end{tabular}
	\label{tab:screencast} \\
	\scriptsize
\end{table}%

\subsubsection{Manual labeling}
\label{sec:manuallabel}
Given an action screencast, the manual labeling has two main steps: video segmentation and structured action description generation.
The screencast is decoded to a sequence of frames at 5fps sampling rate.
We discard frames with no screen changes by computing image similarity because they contain no action information, which removes about 49\% of frames in the raw screencasts.
For efficient and consistent labeling, we develop a web application by which the annotator can view and navigate a screencast by frames, with the change regions highlighted.
The annotators use this tool to annotate the start and end frame of an action fragment and to select command/widget class (as defined in Section~\ref{sec:define}) and enter location phrase for the action fragment.
Location is a free-form phrase and should be described specifically as ``in toolbar'', ``in popup menu'', etc., rather than general window areas such as ``at top'' or ``on the right screen'', because specific location information would be more human understandable.

Two authors participate in the label processing.
The two annotators use the five studied applications regularly in their work and are very familiar with the GUIs and user operations of these applications.
Following the common practice for action video labeling~\cite{zhao2019actionnet}, two annotators label the whole dataset independently.
When there are disagreements, two annotators discuss to decide the final labels.
The labeling process took about 7 person-weeks.
Table~\ref{tab:overallstatistics} summarizes the statistics of labeled action fragments. 
The five applications have similar numbers of action fragments.
We obtain in total 7260 action fragments.
The action fragment duration is about 1 second (5 frames) on average.
Each user action is described by about 4 words including command name, widget name and location phrase.
The location vocabulary contains 71 distinct words.

\subsection{Evaluation metrics}
We evaluate command and widget classification by Precision, Recall, F1-score and Accuracy.
The correctness of a prediction is determined against the human label (i.e. ground truth) for an input action fragment.
Accuracy is an overall performance of 11 commands (or widgets), which is computed by the number of action fragments predicted correctly over all action fragments in the test dataset.  
As location phrase generation is a video captioning task, we evaluate it by BLEU~\cite{papineni2002bleu}, ROUGE~\cite{lin2004rouge}, METEOR~\cite{banerjee2005meteor} and CIDEr~\cite{vedantam2015cider}, which are widely used in image captioning tasks~\cite{chen2015microsoft}.
BLEU, ROUGE and METEOR roughly correspond to precision, recall and F1 in the classification task.
CIDEr uses TF-IDF to weight the words in the phrase.
In this work, considering the short length of location phrases (2.16$\pm$0.42), we compute 1-gram BLEU, ROUGE, METEOR and CIDEr.

\begin{table}
	\centering
	\caption{Overall statistics of labeled action dataset}
	\begin{tabular}{|c|c|c|c|c|}
		\hline
		\textbf{App} &\textbf{\#Action} & \textbf{\tabincell{c}{Duration (s) \\ (mean $\pm$ std)}} & \textbf{\tabincell{c}{\#Words \\ (mean $\pm$ std)}} & \textbf{VocSize} \\
		\hline
		PS    & 1629 & 0.93 $\pm$ 0.75 & 4.43 $\pm$ 0.65 & 42 \\
		\hline
		WinSet  & 1524 & 1.42 $\pm$ 1.13 & 4.40 $\pm$ 0.60 & 28 \\
		\hline
		Firefox      & 1425 & 0.86 $\pm$ 1.06 & 4.16 $\pm$ 0.43 & 42 \\
		\hline
		Zoom         & 1352 & 0.98 $\pm$ 1.01 & 4.10 $\pm$ 0.31 & 33 \\
		\hline
		Word         & 1330 & 1.29 $\pm$ 1.32 & 4.03 $\pm$ 0.18 & 40 \\
		\hline
		\textbf{Total}        & \textbf{7260} & \textbf{0.99 $\pm$ 1.10} & \textbf{4.16 $\pm$ 0.42} & \textbf{71}\\
		\hline
	\end{tabular}
	\label{tab:overallstatistics} \\
	\scriptsize
\end{table}%

\section{Evaluation Results and Findings}
\label{sec:results}

We conduct extensive experiments on our labeled action dataset to investigate the following three research questions:
\begin{itemize}
	\item \textbf{RQ1.} What is the overall performance of the proposed model for command classification, widget classification and location phrase generation?
	
	\item \textbf{RQ2.} How do the three input data streams and the joint learning affect the model performance?
	
    \item \textbf{RQ3.} How does the multi-task learning affect the model performance?
    
\end{itemize}

\subsection{Overall model performance (RQ1)}

\subsubsection{Motivation}
Our model generates structured user action descriptions which include 11 command classes, 11 widget classes, and free-form location phrases for user action screencasts.
Different commands, widgets and locations have diverse visual and temporal characteristics and changes the UIs differently.
This RQ aims to investigate the overall performance of our model in this diverse and challenging learning setting.

\begin{table}
	\centering
	\caption{Overall model performance}
	\begin{tabular}{|c|c|c|c|c|}
		\hline
		\textbf{Output} &\textbf{Class} & \textbf{Precision} & \textbf{Recall} & \textbf{F1-score} \\
		\hline
		\multirow{11}{*}{\tabincell{c}{Command \\ Acc = 0.84}} 
		 & Click       & 0.89 & 0.93 & 0.91 \\
		 \cline{2-5}
		 & Type        & 0.91 & 0.88 & 0.90 \\
		 \cline{2-5}
		 & Drag        & 0.93 & 0.84 & 0.88 \\
		 \cline{2-5}
		 & Hover       & 0.85 & 0.89 & 0.87 \\
		 \cline{2-5}
		 & Select      & 0.93 & 0.82 & 0.87 \\
		 \cline{2-5}
		 & Scroll down & 0.85 & 0.83 & 0.84 \\
		 \cline{2-5}
		 & Appear      & 0.79 & 0.76 & 0.78 \\
		 \cline{2-5}
		 & Zoom in     & 0.69 & 0.80 & 0.74 \\
		 \cline{2-5}
		 & Scroll up   & 0.64 & 0.83 & 0.72 \\
		 \cline{2-5}
		 & Disappear   & 0.71 & 0.73 & 0.72 \\
		 \cline{2-5}
		 & Zoom out    & 0.78 & 0.62 & 0.69 \\
		 \cline{2-5}
	     & \textbf{Average}     & \textbf{0.82} & \textbf{0.83} & \textbf{0.82} \\
		\hline
		\multirow{11}{*}{\tabincell{c}{Widget \\ Acc = 0.87}} 
	 	 & Image    & 0.95 & 0.87 & 0.91 \\
	 	 \cline{2-5}
	 	 & Button   & 0.86 & 0.91 & 0.88 \\
	 	 \cline{2-5}
	 	 & Text     & 0.89 & 0.87 & 0.88 \\
	 	 \cline{2-5}
	 	 & Icon     & 0.88 & 0.85 & 0.86 \\
	 	 \cline{2-5}
		 & Checkbox & 1.00 & 0.75 & 0.86 \\
		 \cline{2-5}
		 & Window   & 0.93 & 0.78 & 0.85 \\
		 \cline{2-5}
		 & Others   & 0.85 & 0.83 & 0.84 \\
		 \cline{2-5}
		 & Popup    & 0.78 & 0.87 & 0.83 \\
		 \cline{2-5}
	 	 & Dropdown & 0.83 & 0.81 & 0.82 \\
	 	 \cline{2-5}
		 & Tab      & 0.92 & 0.73 & 0.81 \\
		 \cline{2-5}
		 & Page     & 0.70 & 0.92 & 0.80 \\
		 \cline{2-5}
		 & \textbf{Average}  & \textbf{0.87} & \textbf{0.84} & \textbf{0.85} \\
		\hline
		\multirow{2}{*}{Location} & \textbf{BLEU} & \textbf{ROUGH} & \textbf{METEOR} & \textbf{CIDEr} \\
		\cline{2-5}
		 & 0.77 & 0.76 & 0.51 & 2.85 \\
		\hline
	\end{tabular}
	\label{tab:eachclass} \\
	\scriptsize
\end{table}%

\subsubsection{Method}
In this experiment, we combine the action data of five applications, and split it into 80\% for model training and 20\% for model testing.
We use the full model for this evaluation, which takes three input data streams (i.e. original frames, cropped change regions and similarity maps) and simultaneously predicts three outputs: command class, widget class and location phrase.
The model is trained for 10 epochs until convergence.
We perform 5-fold cross validation and report the average performance metrics.

\subsubsection{Result}
Table~\ref{tab:eachclass}\footnote{Readers can refer to this GitHub \href{https://github.com/DehaiZhao/SeeAction}{https://github.com/DehaiZhao/SeeAction} for all experiment results in this paper.} shows the full model's performance on the three outputs\footnote{We also examine the model performance on each application, which is similar to the overall performance. Please see the results in the \href{https://github.com/DehaiZhao/SeeAction}{Github repo}}.
For command classification, it achieves an average of 0.84 accuracy and 0.82 F1-score.
Both ``Click'' and ``Type'' have higher F1-score than other command classes, but the reasons are not the same.
The large amount of ``Click'' instances in the dataset makes the model well-trained for the ``Click'' class and results in better performance.
Although ``Type'' does not have many instances, it always interacts with text which has similar visual features and is easy to learn.
We find that our model is often confused between ``Scroll down'' and ``Scroll up''. 
In fact, most incorrectly-recognized ``Scroll down'' is classified as ``Scroll up''.
This is because the two actions have very similar visual features, and ``Scroll down'' has better performance because it has more instances than ``Scroll up'' in the dataset.
A Similar situation is observed for ``Zoom in'' versus ``Zoom out''.
``Appear'' and ``Disappear'' have relatively low F1-score, and most incorrectly-classified ``Appear'' and ``Disappear'' instances are predicted as ``Click''.
This is because click often triggers appearing or disappearing of certain UI parts.

\begin{table*}
	\centering
	\caption{Impact of single-, two- and three-input image sequences}
	\begin{tabular}{|c|c|c|c|c||c|c|c|c||c|c|c|c|}
		\hline
		& \multicolumn{4}{|c||}{\textbf{Command}} & \multicolumn{4}{c||}{\textbf{Widget}} & \multicolumn{4}{c|}{\textbf{Location}} \\
		\hline
		\textbf{Input} & \textbf{Prec} & \textbf{Recall} & \textbf{F1} & \textbf{Accu} & \textbf{Prec} & \textbf{Recall} & \textbf{F1} & \textbf{Accu} & \textbf{BLEU} & \textbf{ROUGH} & \textbf{METEOR} & \textbf{CIDEr} \\
		\hline
		CropCR                   & 0.76 & 0.72 & 0.73 & 0.74 & 0.71 & 0.66 & 0.68 & 0.70 & 0.63 & 0.63 & 0.47 & 0.88 \\
		\hline
		Origin                   & 0.65 & 0.62 & 0.63 & 0.67 & 0.68 & 0.65 & 0.66 & 0.66 & 0.63 & 0.63 & 0.47 & 0.89 \\
		\hline
		SimMap                   & 0.70 & 0.67 & 0.68 & 0.73 & 0.80 & 0.76 & 0.77 & 0.77 & 0.62 & 0.62 & 0.46 & 0.87 \\
		\hline
		CropCR + Origin          & 0.75 & 0.73 & 0.74 & 0.76 & 0.73 & 0.74 & 0.73 & 0.75 & 0.66 & 0.66 & 0.48 & 1.40 \\
		\hline
		CropCR + SimMap          & 0.77 & 0.73 & 0.74 & 0.78 & 0.81 & 0.78 & 0.79 & 0.79 & 0.63 & 0.63 & 0.47 & 0.86 \\
		\hline
		Origin + SimMap          & 0.77 & 0.73 & 0.73 & 0.77 & 0.83 & 0.78 & 0.80 & 0.80 & 0.63 & 0.63 & 0.46 & 0.99 \\
		\hline
		CropCR + Origin + SimMap & 0.82 & 0.83 & 0.82 & 0.84 & 0.87 & 0.84 & 0.85 & 0.87 & 0.77 & 0.76 & 0.51 & 2.85 \\
		\hline
	\end{tabular}
	\label{tab:jointlearning} \\
	\scriptsize
\end{table*}%

Widget classification achieves an average of 0.87 accuracy and 0.85 F1-score, and the F1-score for all widget classes is above 80\%.
``Image'' achieves 91\% F1-score, because image-related actions usually involve larger change regions than the actions on small widgets such as button and icon.
Furthermore, images have more salient features than other large widgets such as window and page.
``Button'' and ``Text'' have F1-score of 88\%, because our dataset has a large number of ``Button'' instances, and all ``Text'' widgets have similar visual features.
We find that most incorrectly-classified ``Dropdown'' instances are predicted as ``Button'' or ``Popup''.
This is because ``Dropdown'' has very similar appearance as ``Button'' before it is expanded, and clicking ``Dropdown'' usually triggers a menu appearing, which looks like ``Popup''.
Some ``Window'' instances are incorrectly predicted as ``Popup'' or ``Page'' as these widgets share similar visual features.
In fact, it might not be necessary to distinguish these window-like widgets involved in user actions, such as [zoom-in] [window] or [zoom-out] [page].

\begin{figure}
	\centering
	\includegraphics[width=\linewidth]{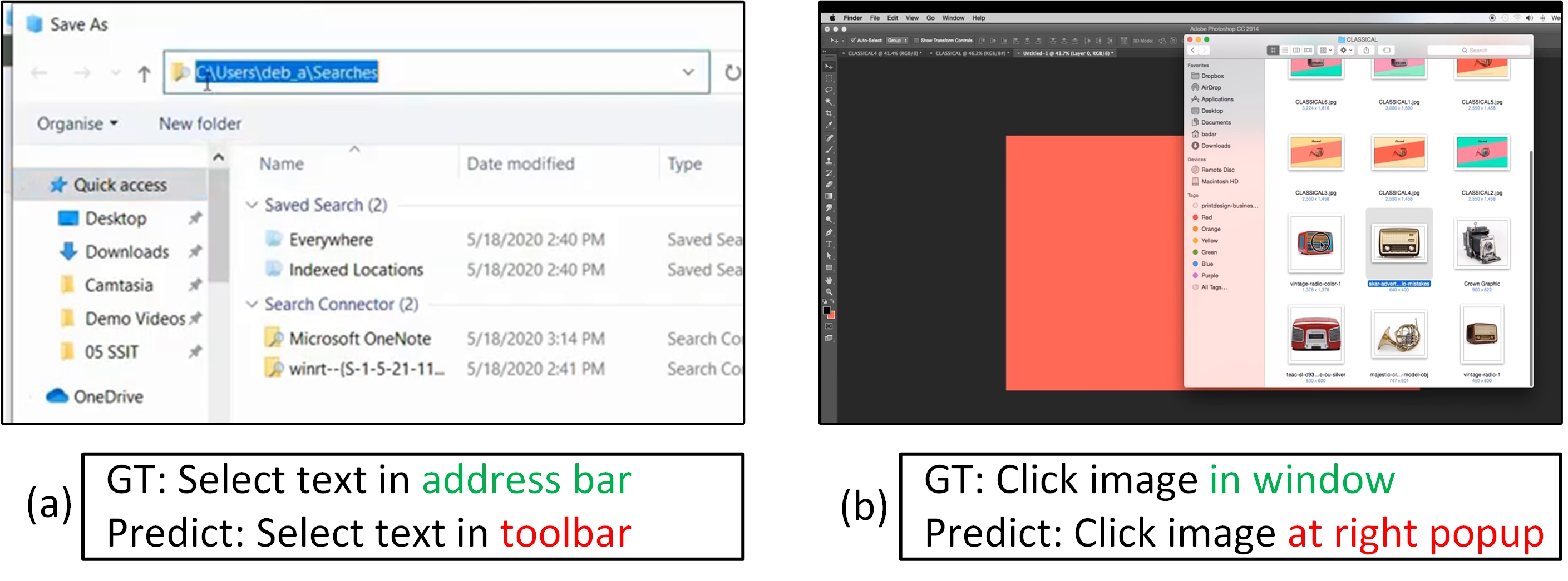}
	\caption{Failure examples of location prediction}
	\label{fig:failure_location}
        \vspace{-1em}
\end{figure}

Location phrase generation achieves 0.77 BLEU, 0.76 ROUGH, 0.51 METEOR and 2.85 CIDEr, which means our model can generate location phrases with good precision, recall and F1 and can generate the most important location information.
Compared with command and widget classification, location phrase generation is a more challenging task because it needs to compose an appropriate free-form phrase for many similar GUI contexts and widgets.
We show two failure examples in Figure~\ref{fig:failure_location}.
In Figure~\ref{fig:failure_location}(a), user selects text in address bar, but the model predicts the location  as ``toolbar''.
In Figure~\ref{fig:failure_location}(b), the model predicts ``at right popup'' while the human label is ``in window''.
The wrong predictions could be caused by the fact that both address bar and toolbar appear at the top of application window, and both window and popup are container widgets and share similar visual features.

\noindent\fbox{\begin{minipage}{8.4cm} \emph{Our model can accurately predict command and widget classes. Most confusions come from the commands and the widgets with similar visual features, such as scroll down/up, dropdown versus button, window versus page. Location phrase generation is a more challenging task, but the generation results by our model are acceptable.} \end{minipage}}\\

\subsection{Impact of joint learning (RQ2)}
\subsubsection{Motivation}
Our model takes three input image sequences: original frames, cropped change regions, and similarity maps, which are designed to play complementary roles to represent user action context and features.
In this RQ, we investigate the impact and synergy of the three input image sequences by ablation study.

\subsubsection{Method}
We develop six variants of input: three single-input variants (cropped change regions (CropCR), originial frames (Origin) or similarity maps (SimMap)), three two-inputs variants (Crop-CR+Origin, CropCR+SimMP, or Origin+SimMap).
These variants use the same model configuration as the full model but ablate the corresponding sub-model(s) for the absent input sequences.
The command classifier, widget classifier and location generator remain unchanged.
We use the same data for training model variants as for the full model (Origin+CropCR +SimMap).
All model variants are trained for 10 epochs until convergence.
We perform 5-fold cross validation and report the average performance metrics.

\begin{table*}
	\centering
	\caption{Performance of multi-task learning versus independent learning or traditional video captioning (UnSentGen)}
	\begin{tabular}{|c|c|c|c|c||c|c|c|c||c|c|c|c|}
		\hline
		& \multicolumn{4}{|c||}{\textbf{Command}} & \multicolumn{4}{c||}{\textbf{Widget}} & \multicolumn{4}{c|}{\textbf{Location}} \\
		\hline
		\textbf{Input} & \textbf{Prec} & \textbf{Recall} & \textbf{F1} & \textbf{Accu} & \textbf{Prec} & \textbf{Recall} & \textbf{F1} & \textbf{Accu} & \textbf{BLEU} & \textbf{ROUGH} & \textbf{METEOR} & \textbf{CIDEr} \\
		\hline
		Independent    & 0.83 & 0.80 & 0.81 & 0.83 & 0.85 & 0.79 & 0.81 & 0.83 & 0.66 & 0.66 & 0.49 & 1.46 \\
		\hline
		UnSentGen      & 0.32 & 0.30 & 0.30 & 0.50 & 0.19 & 0.17 & 0.17 & 0.37 & 0.59 & 0.59 & 0.45 & 0.80 \\
		\hline
		Our model & 0.82 & 0.83 & 0.82 & 0.84 & 0.87 & 0.84 & 0.85 & 0.87 & 0.77 & 0.76 & 0.51 & 2.85 \\
		\hline
	\end{tabular}
	\label{tab:independent} \\
\end{table*}%

\subsubsection{Result}

Table~\ref{tab:jointlearning} reports the performance of the six input variants and the full model.
With single-input sequence, the model achieves only about 0.7 F1-score and accuracy for predicting commands and widgets.
The cropped change region is good at predicting commands, with 74\% accuracy and 73\% F1-score.
Similarity map has better performance than original frame, especially for predicting widgets.
The grayscale similarity map contains much less details than the original RGB frame, but this excludes much noise irrelevant to the action which is beneficial.
In addition, the similarity map can display accurate contour of a change region instead of a rectangle bonding box, which is important for recognizing widgets.
For example, ``Button'' usually has more smooth edge than ``Text'', and ``Checkbox'' is a small rectangle in general.
In contrast, original frame contains too many details, which often blur the most important visual features related to user actions, especially for those commands or widgets involving small change regions, for example, type a short word.

Adding one more input sequence can improve the overall model performance, especially adding cropped change regions or similarity maps to original frames.
For example, original-frame-only-input has poor performance for recognizing small widgets such as ``Button'', ``Checkbox'' and ``Icon'', but adding similarity map can significantly improve the performance (+0.14 in F1-score and +0.14 in accuracy), because similarity map indicates which region on the original frame the model should pay more attention to.
Furthermore, two-inputs model variants achieve very close performance and the gap between the best and the worst model variants narrow when using two input sequences.
With the use of three input sequences (i.e., full model), we observe further 0.06-0.08 increase in F1-score and accuracy for predicting commands and widgets.

\noindent\fbox{\begin{minipage}{8.4cm} \emph{Each input image sequence has its contributions for recognizing user actions. Combining all three input streams and formulating them as joint learning can improve the overall model performance.} \end{minipage}}\\

\subsection{Impact of multi-task learning (RQ3)}
\subsubsection{Motivation}
Our model generates a structured natural language description in the form of [command] [widget] [location].
Considering the potential latent associations between commands, widgets and locations, we train the model in a multi-task learning setting.
In this RQ, we want to compare our method with independent learning and traditional video captioning. 


\subsubsection{Method}
For independent learning, we optimize the command classifier, the widget classifier and the location phrase generator separately, rather than optimizing the combined loss of the three outputs.
For traditional video captioning~\cite{vinyals2015show}, we train an LSTM decoder to generate an unstructured sentence (e.g., ``click button in popup'') for an action screencast, rather than separate command, widget and location.
In both variant settings, the encoder remains the same as the original model.
The variant models are trained in the same way as the original model.
We perform 5-fold cross validation and report the average performance metrics.


\subsubsection{Result}
Table~\ref{tab:independent} presents the performance of the two variant settings and the original model.
Independent learning results in marginal decrease in F1-score and accuracy for predicting commands and 0.04 decrease in F1-score and accuracy for predicting widgets, compared with multi-task learning.
For location phrase generation, independent learning results in relatively larger performance degradation (-0.09 BLEU, -0.06 ROUGE, -0.02 METEOR and -1.39 CIDEr), compared with multi-task learning.
This result agrees with the nature of multi-task learning.
It is an inductive transfer mechanism whose principle goal is to improve generalization performance while keeping the model prediction performance~\cite{caruana1997multitask}.
To gain the global optimization of all subtasks, tradeoffs would be made on the performance of some subtasks.
In our work, multi-task learning keeps the performance of command and widget classification, while the more challenging location phrase generation task benefits from the command and widget classification in the multi-task learning setting.

\begin{figure}
	\centering
	\includegraphics[width=\linewidth]{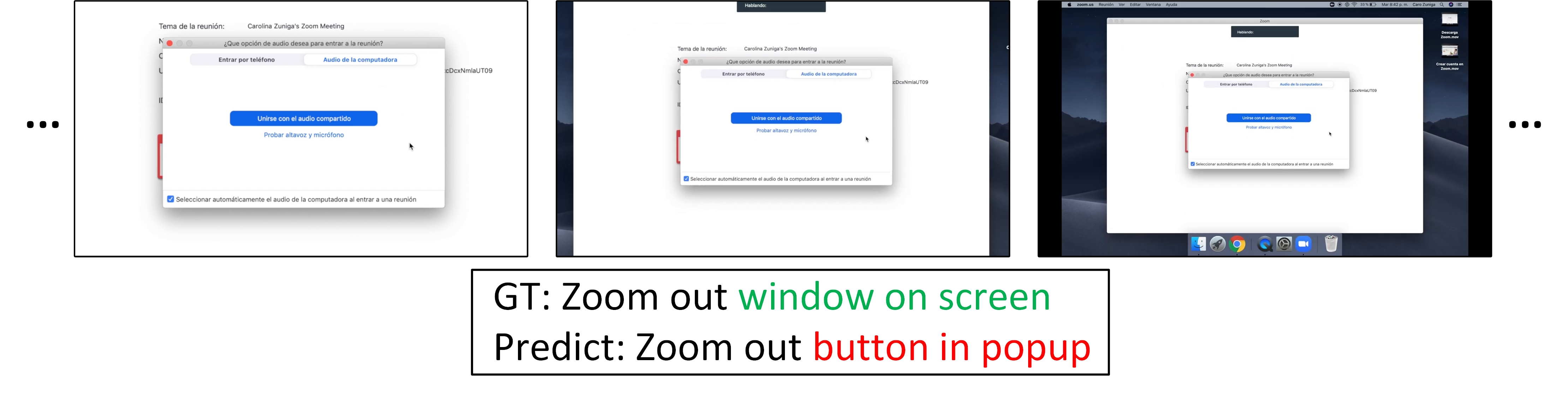}
	\caption{Failure example of traditional video captioning}
	\label{fig:failure_lstm}
\end{figure}

\begin{figure*}
	\centering
	\includegraphics[width=\textwidth]{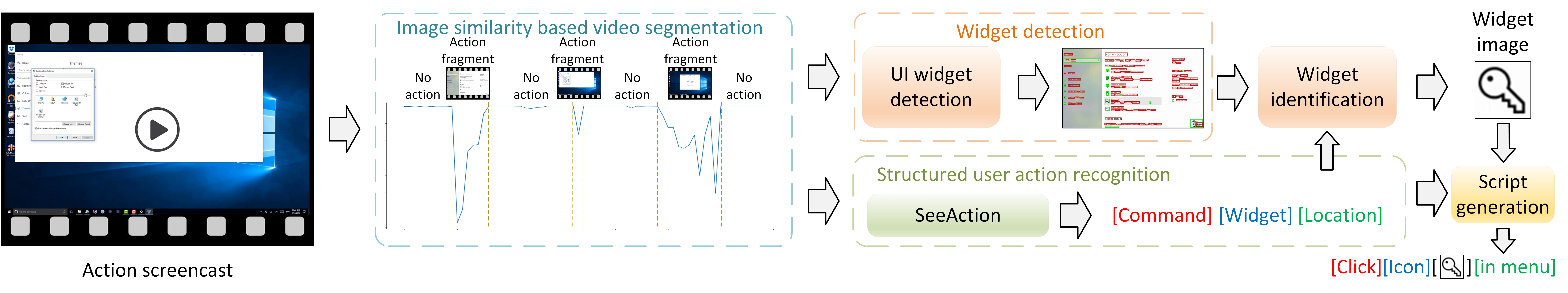}
	\caption{Our Screencast-to-ActionScript Tool}
	\label{fig:pipeline}
\end{figure*}

For traditional video captioning (UnSentGen row), we observe many strange outputs such as ``zoom out button in popup'' in Figure~\ref{fig:failure_lstm}, which is an impossible combination of command and widget.
Such outputs result in very low F1-score and accuracy for predicting commands and widgets.
This is because command and widget classification often require different spatio-visual features to make accurate prediction.
In our structured prediction, two separate MLPs are used for command and widget classification respectively, which can extract command- or widget-specific features.
In contrast, the unstructured sentence generation by traditional video captioning relies on one common feature vector for generating all three types of action information, which is challenging to satisfy all at the same time.
Furthermore, the unstructured sentence generation has to generate a longer sentence from a larger vocabulary (combining command labels, widget labels and location words) than the three separate predictions.
This inevitably increases the task difficulty.

\noindent\fbox{\begin{minipage}{8.4cm} \emph{Multi-task learning slightly improves the performance of command and widget classification, but it benefits the more challenging location phrase generation task. Traditional video captioning cannot reliably generate an unstructured sentence which contains all necessary command, widget and location information.} \end{minipage}}\\

\section{Pilot Study: Bug Reproduction}

Having evaluated our model's performance for structured user action recognition from screencasts, we would like to demonstrate a practical application on bug reproduction that our model enables.

\subsection{Motivation}
Bug reproduction is an essential software engineering task~\cite{chaparro2019assessing, zhao2019recdroid, moran2016automatically, fazzini2018automatically}.
There are two common ways to report a bug: text description and screencast demonstration.
The study~\cite{chaparro2019assessing} shows that many users are not professional enough to write a high quality bug report with clear and complete steps to reproduce.
This not only increases the burden of software maintainers to understand and analyze the bugs, but also creates a barrier to automated bug reproduction techniques based on the steps-to-produce descriptions in bug reports~\cite{zhao2019recdroid, moran2016automatically}.
In contrast, recording screencasts lowers the bar for ordinary users to report bugs and it can cover every single detail of a bug (when and how it occurs).
GitHub has announced that video upload is available on its platform~\cite{github}, meaning that video data is becoming more and more important in software community.
However, the image nature of screencasts makes it hard to integrate video content with existing text-based tools.
In this work, we conduct a pilot study of this potential usage.
In particular, we investigate if the generated human-understandable action scripts from bug screencasts can help people reproduce the bugs. 

\subsection{A Screencast-to-ActionScript Tool}
We develop a screencast-to-actionscript (S2AS) tool for our pilot study.
As shown in Figure~\ref{fig:pipeline}, S2AS integrates our SeeAction model with a heuristic-based video segmentation method and a UI widget detection method, thus constituting a full reverse-engineering process that converts an input action screencast into a script of structured user actions, such as those shown in Figure~\ref{fig:structured}.

Based on our video labeling experience and the observation of our labeled action fragments, we design a simple image similarity based method for video segmentation, driven by the fact that when a user action occurs on the UI, there will be more or less pixel changes.
As illustrated in Figure~\ref{fig:pipeline}, for an action screencast in our dataset, user actions result in low screen similarities during the action time periods, while the periods without user actions have high screen similarities and remain stable over time.
Based on this observation, S2AS segments the screencast into a series of action fragments by identifying the time points when the screen similarities start and finish changing.

For widget detection, we use an existing technique UIED~\cite{xie2020uied} which can detect the bounding box of a widget and recognize its class.
For detected text widgets, UIED uses Google OCR~\cite{ocr} to convert text widget images to texts.
S2AS applies UIED to an action fragment in the reverse order (from the last frame to the first).
UIED detects all widgets on a UI image.
If UIED detects a widget on a frame whose bounding box overlaps with the change region identify by SeeAction for that frame and that has the same class as the widget class predicted by SeeAction, S2AS identifies this widget as the target widget of the user action.
If the target widget is a non-text widget, S2AS crops the widget image by its bounding box.
For text target widget, it takes the OCRed text as the widget information.
Finally, it outputs a complete structured user action based on the structured user action predicted by SeeAction and the identified widget information, such as [click] [Icon] [\img{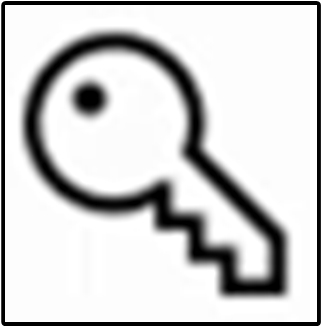}] [in menu].
S2AS applies SeeAction and UIED to all action fragments and generates a script of user actions for the input screencast.
\subsection{Pilot Study}
\subsubsection{Bug Reproduction Dataset}
We randomly sample 100 bug reports with bug screencasts on Firefox Bugzilla~\cite{Bugzilla}.
These bugs were reported during January 2022 - January 2024 and have no overlap with our labeled dataset.
They are related to a wide range of Firefox functionalities (e.g., website browsing, web page inspection, menu bar customization).
By inspecting the text description of the collected bug reports, we summarize them to five quality grades, which include High Quality (HQ), Low Quality - Ambiguous Step (LQ-AS), Low Quality - Vocabulary Mismatch (LQ-VM), Low Quality - Missing Step (LQ-MS), and No Steps to Reproduce (N-S2R).
High quality means the bug reports have detailed, clear and complete steps to reproduce the bugs.
N-S2R means the complete absence of S2R description in bug report.
The three low quality grades are defined according to the S2R quality issues revealed in~\cite{chaparro2019assessing}.
Ambiguous step means a step matches more than one GUI component or event.
Vocabulary mismatch means a step does not match any application interaction.
Missing step means a required step is not described in the bug report.
A bug report may suffer from several low quality issues.
We annotate the most significant issue that prevents the bug reproduction as a bug's quality grade.

\begin{table}
	\centering
	\caption{Bug report quality grades and reproduction results}
	\begin{tabular}{|c|c|c|c|c|c|c|}
		\hline
		\textbf{Q-Grade} &\textbf{HQ} & \textbf{LQ-AS} & \textbf{LQ-VM} & \textbf{LQ-MS} & \textbf{N-S2R} & \textbf{ALL} \\
		\hline
		\#S2R-QG      & 23 & 9 & 10 & 27 & 31 & 100 \\
		\hline
		ReproS2R      & 23 & 3 & 4  & 8  & 0  & 38  \\
		\hline
		ReproGAS      & 15 & 5 & 6  & 16 & 19 & 61  \\
		\hline
	\end{tabular}
	\label{tab:tool} \\
	\scriptsize {Q-Grade: quality grade, \#S2R-QG: the number of bug reports that have different quality grades of Steps to Reproduce, ReproS2R: the number of bug reports that are correctly reproduced by the Steps to Reproduce descriptions, ReproGAS: the number of bug reports that are correctly reproduced by Generated Action Scripts.}
\end{table}%

Two authors annotate the bug report qualities independently, and their initial annotations have 0.81 inter-rater agreement by Cohen's kappa~\cite{cohen1960coefficient}.
For disagreements, they discuss to decide the final grade.
As shown in Table~\ref{tab:tool}, only 23 out of 100 bug reports have high-quality S2R descriptions, and 31 bug reports do not provide S2R description at all (the bug report may simply leave a statement ``see the attached video'').
For the rest 46 low-quality bug reports, 27, 10 and 9 of them have significant issues in missing steps, vocabulary mismatch and ambiguous steps, respectively.
Our results are similar to the observation of S2R quality issues in~\cite{chaparro2019assessing}.

\subsubsection{Study Procedure}
We recruit six graduate students from our school to participate in our study.
Through a pre-study survey, all participants use Firefox regularly and are familiar with the main functionalities involved in the bugs.
The participants do not know any background knowledge of this work before finishing the experiment.
We process the bug screencasts in bug reports by our S2AC tool, and generate an action script for each of them.
We collect the S2R descriptions (if any) in the bug reports.
The six participants are split to two groups.
Three of them (ReproS2R) are given the original S2R descriptions in the bug reports for bug reproduction, while the other three (ReproGAS) use the generated action scripts by our tool.
All participants screen-record their bug reproduction process.
Following the practice to evaluate the success of record-and-replay tools~\cite{guo2019sara, sahin2019randr, qin2016mobiplay, yu2021layout}, we manually check the Externally Visible State (EVS) of the bug reproduction, and regard the reproduction as a success if and only if the EVS by reproduction is the same as the EVS of the reported bug.

\subsubsection{Results and Analysis}

Table~\ref{tab:tool} reports the union of the bug reports that were successfully reproduced by the three participants using the original S2R descriptions (ReproS2R) or the generated action scripts (ReproGAS).
Obviously, unless the ReproS2R participants watch the bug screencasts, it is impossible to reproduce the 31 bugs without the S2R descriptions (N-S2R). 
Among the 69 bug reports with the S2R descriptions, only 38 of them are reproduced successfully by using the original S2R descriptions, and the majority (23) are high quality bug reports.
The participants have much lower success rate (only about 30\%) for reproducing bugs based on low quality S2R descriptions.
For example, given an ambiguous step ``Search mozilla'' (LQ-AS), one can type ``mozilla'' in either search bar or address bar of the browser, or may search some other keywords such as ``what is mozilla''.
Another example is vocabulary mismatch (LQ-VM) ``Disable OMTC''.
None of the three ReproS2R participants know what OMTC (off main thread compositing) is before the experiment and are stuck in this step.
Reproducing the bug reports that miss important steps (LQ-MS) is also a big challenge.
For example, the S2R of the bug report may mention only ``set default search engine as Google''.
The participants have to figure how to reach the particular setting page and change the setting. 

Using the generated action scripts, the number of successfully reproduced bugs increases to 61.
Actually, the overall success rate and the success rates for different quality grades are all about 60\%-65\%, because using the generated action scripts to reproduce the bugs has nothing to do with the presence and quality of the original S2R descriptions.
19 of 31 bugs without the S2R descriptions and about 60\% of bugs with low-quality S2R descriptions are successfully reproduced based on the generated action scripts. 
Compared with the 30\% reproduction success rate based on the original low-quality S2R descriptions, this is a promising result as a tool like S2AC could be leveraged to help ``lazy'' bug reporters write good-quality S2R descriptions.
Of course, compared with the high-quality S2R descriptions written by human (100\% reproduction success), the generated action scripts still have room for improvement.

For the whole screencast-to-actionscript process, we find that video segmentation deserves more research as the inaccurate segmentation will inevitably affect the subsequent action recognition.
For example, the user scrolls up and down the web page back and forth quickly, leading to less accurate segmentation between the consecutive similar user actions.
Some user actions may also be over-segmented, leading to many too fine-grained actions.
For example, a ``click button'' action are segmented into ``hover button'' and ``click button'', and selecting a long text are segmented to multiple ``select text''.
Although much of such less ideal segmentation can still be understood and applied by human participants, some may cause the human confusion and the reproduction failures.
For example, a ``click checkbox'' action is segmented into two (or more) ``click checkbox'' actions, and the second (and following) ``click checkbox'' may lead to unexpected results.
Such less ideal segmentation would also negatively affect the automatic replay tools~\cite{bernal2020translating, yu2021layout, airtest, sikulix}.

For widget detection, we encounter some difficulty of adjusting two important parameters of UIED, $min grade$ and $min ele area$.
Too small value results in many noisy bonding boxes, and too large value misses many small widgets.
After the experiments, we set $min grade = 40$ and $min ele area = 3$ which achieves the best result on our dataset.
But UIED still detects some noisy widgets or misses some widgets, which result in the wrong or incomplete action scripts affecting the bug reproduction.
For example, an icon is misclassified as text, resulting in the mismatch of widget class, and consequently a missing step in the generated action script.

\noindent\fbox{\begin{minipage}{8.4cm} \emph{Our pilot study, albeit by no means conclusive, demonstrates the promise of non-intrusive screencast-to-actionscript reverse-engineering tool for bug reproduction. It also reveals the challenges in building the tool pipeline and the improvement for video segmentation and widget identification which we leave as our future work.} \end{minipage}}\\

\section{Related Work}
\label{sec:relatedwork}

HCI actions can be recorded using instrumentation-based or computer vision based methods.
Instrumentation-based methods are all more or less intrusive, relying on source code or customized OS~\cite{qin2016mobiplay, hu2015versatile}, or OS or GUI framework accessibility support~\cite{kim2008tracking, bao2015activityspace, bao2018vt}, or runtime customization~\cite{guo2019sara, sahin2019randr}.
Computer-vision based methods offer an alternative non-intrusive way, and receive growing attention in recent years.
For examples, ActionNet~\cite{zhao2019actionnet} extracts primitive mouse and keyboard commands between two consecutive frames in an action screencast.
V2S~\cite{bernal2020translating} recognizes tap command in mobile apps usage videos based on tap indicators.
Other works investigate the extraction of GUI widgets (e.g., Waken~\cite{banovic2012waken}, Prefab~\cite{dixon2010prefab}) or application contents (e.g., source code~\cite{bao2017extracting, ponzanelli2016too}).
Our work is the first non-intrusive method to extract command, widget and location as a whole structured action, and our approach does not limit action fragment lengths, nor does it assume special visual indicators. 

Recognizing HCI actions can help the analysis of developer behavior~\cite{bao2015activityspace, bao2017extracting, hilbert2000extracting}.
It is also the foundation for UI automation tasks, for example, record-and-replay testing~\cite{guo2019sara, sahin2019randr, qin2016mobiplay, chen2022extracting} and robotic process automation~\cite{qian2020roscript}.
A wide range of UI automation tools are available, such as Selenium~\cite{selenium}, Appium~\cite{appium}, Robotium~\cite{robotium}, Monkeyrunner~\cite{monkeyrunner}, UIAutomator~\cite{uiautomator}.
All these tools rely on OS or GUI framework accessibility support.
In contrast, visual script based UI automation such as Sikuli~\cite{sikulix}, AirTest~\cite{airtest}, UiPath~\cite{uipath} and UI.Vision~\cite{ui.vision} use computer vision techniques to match widgets to be operated.
Although they are platform independent, widget matching is sensitive to changes of screen size and resolution and widget appearance~\cite{yu2021layout, guo2019sara}.
All these UI automation techniques focus on action replay.
In contrast, our work focuses on non-intrusive action recognition from screencasts.

Our approach can be regarded as a vision-to-text task.
In computer vision community, image captioning~\cite{vinyals2015show, you2016image, johnson2016densecap} and video captioning~\cite{xu2017learning, li2019residual, gao2017video, pan2017video} are two typical vision-to-text tasks.
Image captioning techniques have been adopted for generating GUI code from GUI design images~\cite{beltramelli2018pix2code, chen2018ui} or generating widget labels to improve accessibility~\cite{chen2020unblind, mehralian2021data, li2020widget, li2022spotlight, feng2023read}.
However, these works generate unstructured text for describing the image or video.
Our experiment in RQ3 show this cannot satisfy the need for complex user action recognition.
Video segmentation~\cite{zhao2017temporal, xu2017r}, video retrieval~\cite{yan2024semantic, cooper2021takes} and object detection~\cite{ren2015faster, redmon2016you} are related techniques, which can be integrated with our approach to build the whole screencast-to-actionscript pipeline.
Object detection models have been used for random GUI testing~\cite{white2019improving}, and GUI widget specific detection method has also been proposed~\cite{xie2020uied}.
Other remotely related work includes predicting screens with code~\cite{bergh2020curated}, GUI visual quality~\cite{liu2020owl}, game visual quality~\cite{ye2021empirical} and actionable GUI areas~\cite{yazdanibanafshedaragh2020deep}.
These works make prediction on static GUI images, while our goal is action recognition from screencasts.

\section{Threats to Validity}
One threat to internal validity is the manual labeling bias of the screencast dataset.
In order to minimize the errors, two authors label the whole dataset independently and finally reach agreement by discussion.
We find that the disagreements mainly come from judging location of user action.
For example, two authors give label of ``in search bar'' and ``in textbox'' to a video fragment, but they refer to the same location in Firefox, which has low impact to user action representation.
Another internal threat is from the participants' bias of pilot study.
None of the participant know the knowledge of our work before completing the experiment.

Threats to external validity include the selection of desktop applications and AI models, and widget detection accuracy.
We use 5 popular desktop applications in our work and further evaluation is required on other types of applications such as video game and mobile applications.
In terms of AI models, we applied 3D CNNs and LSTM in this work to support video action recognition. 
However, it is easy to replace those modules with alternative AI models such as Transformer\cite{vaswani2017attention} and BERT\cite{devlin2018bert}
Furthermore, the pilot study involves multiple steps of processing and the error accumulation of each step (especially widget detection by UIED) makes it not perfect enough to support all practical applications.
This work focuses on generating structured natural language description from screencasts and we will try to optimize the pipeline by improving the performance of each step.

\section{Conclusion and Future work}
This work proposes a novel computer vision based model to reverse-engineer structured user actions from only the visual information recorded in screencasts.
Our model extracts the spatio-temporal features jointly from the three complementary data streams, original frames, change regions and similarity maps, and predicts the command, widget and location in a multi-task setting.
We evaluate our model on a large dataset of 7260 video-action pairs from five widely-used applications.
Our results confirms the high accuracy of our model (even in the challenging across application training-testing setting) and the effectiveness of our joint-learning and multi-task learning design.
We develop a prototype screencast-to-actionscript tool based on our model, and demonstrate its effectiveness in generating high-quality, human-understandable action scripts for bug reproduction through a pilot study on 100 Firefox bugs.

SeeAction's functionality extends beyond bug report reproduction. 
Its human-readable and structured HCI action descriptions can support a variety of downstream software engineering tasks, including UI automation and testing, broadening its applicability and potential impact.
In the future, we will investigate more robust video segmentation and GUI widget detection methods to improve the screencast-to-actionscript pipeline for non-intrusive UI automation, and further explore its potential applications in various software engineering tasks.

\bibliographystyle{IEEEtran}
\bibliography{reference}

\end{document}